\documentstyle[12pt]{article}

\def\hybrid{\topmargin -20pt  \oddsidemargin 0pt
      \headheight 0pt   \headsep 0pt
      \textwidth 6.25in 
      \textheight 9.5in 
      \marginparwidth .875in
      \parskip 5pt plus 1pt   \jot = 1.5ex}

\hybrid

\def\o+{\oplus}

\def\lra{\longrightarrow}

\def\beqa{\begin{eqnarray}}
\def\eeqa{\end{eqnarray}}
                       
\newcommand{\un}{\underline}

\newcommand{\la}{\lambda}
\newcommand{\si}{\sigma}

\newcommand{\om}{\omega}

\newcommand{\resetcounter}{\setcounter{equation}{0}}

\begin{document}
\thispagestyle{empty}
\rightline{LMU-ASC {73}/10}
\vspace{2truecm}
\centerline{\bf \LARGE On possible Chern Classes of stable
Bundles}
\vspace{.3truecm}
\centerline{\bf \LARGE on Calabi-Yau threefolds}

\vspace{1.5truecm}
\centerline{Bj\"orn Andreas\footnote{andreas@math.hu-berlin.de}$^\$$ 
and Gottfried Curio\footnote{gottfried.curio@physik.uni-muenchen.de; 
supported by DFG grant CU 191/1-1}$^{\scriptsize\mbox{\pounds}}$} 

\vspace{.6truecm}

\centerline{$^\$${\em Institut f\"ur Mathematik}}
\centerline{{\em Humboldt-Universit\"at zu Berlin}}
\centerline{{\em Rudower Chaussee 25, 12489 Berlin, Germany}}

\vspace{.3truecm}
\centerline{$^{\scriptsize\mbox{\pounds}}${\em Arnold-Sommerfeld-Center 
for Theoretical Physics}}
\centerline{{\em Department f\"ur Physik, 
Ludwig-Maximilians-Universit\"at M\"unchen}}
\centerline{{\em Theresienstr. 37, 80333 M\"unchen, Germany}}

\vspace{1.0truecm}

\begin{abstract}
Supersymmetric heterotic string models, built from a Calabi-Yau threefold $X$ 
endowed with a stable vector bundle $V$, usually lead to an anomaly mismatch 
between $c_2(V)$ and $c_2(X)$; this leads to the question whether the 
difference can be realized by a further bundle in the hidden sector.
In math.AG/0604597 a conjecture is stated
which gives sufficient conditions on cohomology classes on $X$ 
to be realized as the Chern classes of a stable reflexive sheaf $V$; 
a weak version of this conjecture predicts the existence of such a $V$ 
if $c_2(V)$ is of a certain form.
In this note we 
prove that on elliptically fibered $X$ infinitely many
cohomology classes $c\in H^4(X, {\bf Z})$ exist which are of this form 
and for each of them a stable $SU(n)$ vector bundle with $c=c_2(V)$ exists.
\end{abstract}

\newpage

\section{Introduction}

To get $N=1$ heterotic string models in four dimensions
one compactifies the tendimensional heterotic string on a Calabi-Yau
threefold $X$ which is furthermore endowed with a polystable 
holomorphic vector bundle $V'$. Usually one takes $V'=(V, V_{hid})$ with
$V$ a stable bundle considered to be embedded in (the visible) $E_8$ 
($V_{hid}$ plays the corresponding role for the second hidden $E_8$); the
commutator of $V$ gives the unbroken gauge group in four dimensions.

The most important invariants of $V$ are its Chern classes
$c_i(V)$, $i=0,1,2,3$. We consider in this note bundles with
$c_0(V)=rk(V)=n$ and $c_1(V)=0$; more specifically we will consider
$SU(n)$ bundles. The net number of generations of chiral particle multiplets
in the fourdimensional effective theory
is given by $N_{gen}(V)=c_3(V)/2$. On the other hand the second Chern class
is important to assure anomaly freedom of the construction: this is encoded
in the integrability condition for the existence 
of a solution to the anomaly cancellation equation
\beqa
\label{anom}
c_2(X)&=&c_2(V)+W.
\eeqa
Here $W$, as it stands, has just the meaning to indicate a possible mismatch
for a certain bundle $V$; it can be understood either as the cohomology class
of (the compact part of the world-volume of) a fivebrane, or as second Chern
class of a further stable bundle $V_{hid}$ in the hidden sector. Furthermore in
the first case the class of $W$ has to be effective for supersymmetry
to be preserved. 

Often one will argue just from the data provided by the Chern classes,
say to secure a certain phenomenologically favored generation number,
and so has to make sure that a corresponding $SU(n)$ bundle 
with suitably prescribed Chern class $c_3(V)$ actually exists.
On the other hand, following the route to solve (\ref{anom}) described,
one has the same problem
for $c_2(V_{hid})=W:=c_2(X)-c_2(V_{vis})$ concerning the hidden bundle.

In [\ref{AFG}] it has been shown that whenever the topological constraint 
can be satisfied with $W=0$ then $X$ and $V$ can be 
deformed to a solution of the anomaly equation even already on the level 
of differential forms (a solution to the system involving the three-form
field-strength $H$, investigated first in [\ref{Strom}], exists). 

This leads to the general question to give sufficient conditions for the
existence of stable bundles with prescribed Chern classes $c_2(V)$
and $c_3(V)$. Concerning this the following conjecture is put forward
in [\ref{DRY}] by Douglas, Reinbacher and Yau (DRY) (actually we use the 
particular case of the conjecture with $c_1(V)=0$).

{\bf DRY-Conjecture.} {\em On a Calabi-Yau threefold $X$ of $\pi_1(X)=0$ a 
stable reflexive sheaf $V$ of rank $n$ and $c_1(V)=0$ 
with prescribed Chern classes $c_2(V)$
and $c_3(V)$ will exist if, for an ample class $H\in H^2(X, {\bf R})$, 
these can be written as (where $C:=16\sqrt{2}/3$)}
\beqa\label{dry}
c_2(V)&=&n\Bigg( H^2+\frac{c_2(X)}{24}\Bigg)\\
c_3(V)&<& C \; nH^3.
\eeqa
Note that the conjecture just predicts 
the existence of a stable reflexive sheaf;
in our examples below $V$ will be a vector bundle.

We will also formulate a weaker version of the conjecture, which is 
implied by the proper (strong) form and concentrates on the
existence of $V$ given that just its (potential) $c_2(V)$ fulfills
the relevant condition.
To refer more easily to the notions involved,
we make first the following definitions.
We restrict to the case of $V$ being 
a vector bundle. 
We will consider rank $n$ bundles of $c_1(V)=0$
and treat actually the case of $SU(n)$ vector bundles.

{\bf Definition.} Let $X$ be a Calabi-Yau threefold of $\pi_1(X)=0$ and
$c\in H^4(X, {\bf Z})$, \\
i) $c$ is called a {\em Chern class} 
if a stable $SU(n)$ vector bundle $V$ on $X$ exists with $c=c_2(V)$\\
ii) $c$ is called a {\em DRY class} if an ample class 
$H\in H^2(X, {\bf R})$ exists (and an integer $n$) with 
\beqa
\label{weak dry}
c_2(V)&=&n\Bigg( H^2+\frac{c_2(X)}{24}\Bigg).
\eeqa

With these definitions we can now state the weak DRY conjecture,
in the framework as we will use it, as follows:

{\bf Weak DRY-Conjecture.} {\em On a Calabi-Yau threefold $X$ of $\pi_1(X)=0$ 
every DRY class $c\in H^4(X, {\bf Z})$ is a Chern class.}

Here it is understood that the integer $n$ occurring in the two
definitions is the same.

The paper has three parts.
In {\em section 2} we give criteria for a class to be a DRY class.
In {\em section 3} we present some bundle constructions
and show that their $c_2(V)$ fulfill these criteria for infinitely many $V$.

\section{DRY classes on elliptic Calabi-Yau threefolds}

\resetcounter

To test these conjectures we choose $X$ to be elliptically fibered over
the base surface $B$
with section $\sigma:B\lra X$ (we will also denote by $\si$ 
the embedded subvariety
$\si(B)\subset X$ and its cohomology class in $H^2(X, {\bf Z})$), 
a case particularly well studied. The typical examples for $B$ 
are rational surfaces like a Hirzebruch surface ${\bf F_k}$
(where we consider the following cases $k=0,1,2$ as only 
for these bases exists a smooth elliptic $X$ with Weierstrass model), 
a del Pezzo surface ${\bf dP_k}$ ($k=0, \dots, 8$)
or the Enriques surface (or suitable blow-ups of these cases).
We will consider specifically bases $B$ for which $c_1:=c_1(B)$ is ample.
This excludes in particular the Enriques surface and the Hirzebruch 
surface\footnote{as $c_1\cdot b = (2b+4f)\cdot b=0$, using here
the notations from footn.~\ref{basepoint}} ${\bf F_2}$.
(The classes $c_1^2$ and $c_2:=c_2(B)$ will be considered as (integral) 
numbers.)

On the elliptic Calabi-Yau space $X$ one has 
according to the general decomposition
$H^4(X, {\bf Z})\cong H^2(X, {\bf Z})\si \oplus H^4(B, {\bf Z})$
the decompositions (with $\phi, \rho \in H^2(X, {\bf Z})$)
\beqa
c_2(V)&=&\phi \si + \om\\
c_2(X)&=&12c_1\si +c_2+11c_1^2
\eeqa
where $\om$ is understood as an integral number 
(pullbacks from $B$ to $X$ will be usually suppressed).

One now solves for $H=a \si + \rho$ (using 
the decomposition $H^2(X, {\bf Z})\cong {\bf Z}\si \oplus H^2(B, {\bf Z})$), 
given an arbitrary but fixed class
$c=\phi \si + \omega \in H^4(X, {\bf Z})$, and has then to check that 
$H$ is ample.
The conditions for $H$ to be ample are (cf.~appendix)
\beqa
H\ \ \ {\rm ample} \;\;\;\;\; &\Longleftrightarrow& \;\;\;\;\; 
a>0, \;\;\;\; \rho-ac_1 \;\; {\rm ample}.
\eeqa 
Inserting $H$ into equ.~(\ref{weak dry}) one gets
the following relations 
\beqa\label{data1}
\phi&=& 2an\rho+n\Big(\frac{1}{2}-a^2\Big)c_1\\
\label{data2}
\frac{1}{n}\omega&=&\rho^2+\frac{5}{12}c_1^2+\frac{1}{2}
\eeqa
(note that $\sigma^2=-c_1\sigma$, cf.~[\ref{FMW}])
where in (\ref{data2})  Noethers relation $c_2+c_1^2=12$ 
for the rational surface $B$ has been used.
This implies 
\beqa 
\rho&=&\frac{1}{2an}\Bigg(\phi-n(\frac{1}{2}-a^2)c_1\Bigg)
\eeqa
As we assumed that $c_1$ is ample one finds that the condition that the class
\beqa
\label{rho condition}
\label{rho - a c1}
\rho-ac_1&=&\frac{1}{2an}\Bigg(\phi-n(\frac{1}{2}+a^2)c_1\Bigg)
\eeqa
is ample leads also to an upper bound on the positive real number $a^2$ 
\beqa
\label{bound}
0\; < &a^2 & < \; b
\eeqa
This bound will be specified in an example below explicitly. 
Furthermore equ.~(\ref{rho condition}) shows that $\phi-\frac{n}{2}c_1$
must necessarily be ample.

Having solved by now (\ref{data1}) in terms of $\rho$ 
one now has to solve the following equation
\beqa\label{omega2}
\frac{1}{n}\omega&=&
\frac{1}{4a^2n^2}\Big(\phi-n(\frac{1}{2}-a^2)c_1\Big)^2
+\frac{5}{12}c_1^2+\frac{1}{2}
\eeqa
in terms of $a$.
Actually the only non-trivial point will be that $a$ is real and satisfies
(\ref{bound}). 
Concretely one gets a quadratic equation $a^4+pa^2+q=0$ in $a^2$ with
\beqa
p&=&-\frac{4}{c_1^2}
\Big(\frac{1}{n}\omega-r\Big)\\
q&=&\frac{4}{(c_1^2)^2}s^2
\eeqa
where we used the abbreviations 
\beqa
r&:=&\frac{1}{2n}\phi c_1 + \frac{1}{6}c_1^2+\frac{1}{2}\\
s&:=&\frac{1}{2n}\sqrt{c_1^2}\sqrt{\Big(\phi-\frac{n}{2}c_1\Big)^2}
\eeqa
Now one has three conditions which have to be satisfied by 
at least one solution $a^2_*$ of this equation, 
namely\footnote{here it is understood that ii) assumes i) satisfied 
and iii) assumes i) and ii) satisfied}
\beqa
i)\;\;\;\;\;\;\;\; \;\;\;\;\;\;\;\;\;\;\;
a^2_*\in {\bf R}  &\Longleftrightarrow & \;\;\;\;\;\;\; p^2\geq 4q 
\;\; \Longleftrightarrow \;\; \Big(\frac{1}{n}\omega-r\Big)^2\geq s^2\\
ii) \;\;\;\;\;\;\;\;\;\;\;\;\;\;\;\;\;\;\;
a^2_*>0 &\Longleftrightarrow &\;\;\;\;\;\; -p> 0
\;\;\; \Longleftrightarrow \;\;\; \frac{1}{n}\omega > r \;\;\\
iii) \;\;\;\;\;\;\;\; \;\;\;\;\;\;\;\;
a^2_*\; \leq \; b &\Longleftrightarrow &\;\; 
\left\{ \begin{array}{lll}
-p< b+\frac{q}{b}\;\;\;\;\;\;\;\;\;\;\;\;\;\;\;\;\;\;\;
\mbox{for} \;\; +\sqrt{} \; \mbox{and} \; b\geq -\frac{p}{2}\\
\;\;\mbox{arbitrary} \;\;\;\;\;\;\;\;\;\;\;\;\;\;\;\;\;\;\;\;
\mbox{for} \;\; -\sqrt{} \; \mbox{and} \; b\geq -\frac{p}{2}\;\;\;\;\;\;\;\;\\
-p> b+\frac{q}{b}\;\;\;\;\;\;\;\;\;\;\;\;\;\;\;\;\;\;\;
\mbox{for} \;\; -\sqrt{} \; \mbox{and} \; b< -\frac{p}{2}
\end{array} \right.
\eeqa

Concerning ii) 
note that necessarily $q> 0$, cf.~the remark after (\ref{bound});
furthermore the evaluation of the condition is independent of the question
which sign for the square root is taken.
Note that in iii) the case where $+\sqrt{}$ is taken and $b<-\frac{p}{2}$
is excluded.

Concerning condition iii) note that 
for $b\geq -p/2$ one gets no further restriction and has to pose {\em in total }
just the first two conditions, i.e. $\frac{1}{n}\omega\geq r+s$. 
By contrast for $b<-p/2$ the condition $-p> b+ \frac{q}{b}$, equivalently 
$\frac{1}{n}\omega> \omega_0(\phi; b):=r+\frac{c_1^2}{4}(b+\frac{q}{b})$,
implies i) and ii). 

As $p$ (and thus $\omega$) occurs in the domain restrictions on $b$ 
one has to rewrite these conditions slightly.
Consider first the regime $b<-\frac{p}{2}$, or explicitly
$\frac{1}{n}\omega>r+\frac{b}{2}c_1^2$, and distuinguish two cases:
the ensuing condition $\frac{1}{n}\omega> \omega_0(\phi; b)$ 
makes sense as an {\em additional} condition (required 
besides the domain restriction $\frac{1}{n}\omega> r+\frac{b}{2}c_1^2$)
only if $r+\frac{b}{2}c_1^2< \omega_0(\phi; b)$, i.e.~for $b< \sqrt{q}$;
on the other hand for $b\geq \sqrt{q}$ one just has to demand 
$\frac{1}{n}\omega> r+\frac{b}{2}c_1^2$.

In the second regime $b\geq -p/2$, or equivalently 
$\frac{1}{n}\omega \leq r+\frac{b}{2}c_1^2$, 
one has the condition $\frac{1}{n}\omega \geq r+s$
(note that these two conditions are compatible just for 
$s\leq \frac{b}{2}c_1^2$, i.e. $b\geq \sqrt{q}$).
Thus in total one can make an $\omega$-independent regime distinction for
$b$ according to $b< \sqrt{q}$ (with the demand $\frac{1}{n}\omega > \omega_0$)
or $b\geq \sqrt{q}$ 
(where one should have either $\frac{1}{n}\omega>r+\frac{b}{2}c_1^2$
or $\frac{1}{n}\omega\leq r+\frac{b}{2}c_1^2$ 
but in that latter case one has to demand
$\frac{1}{n}\omega\geq r+s$; but, 
as $r+s\leq r+\frac{b}{2}c_1^2$ in the present $b$-regime,
one just has to demand that $\frac{1}{n}\omega\geq r+s$).

Note in this connection that, for $\phi$ held fixed, 
$\omega_0(\phi; b)$ becomes large for small and
large $b$ and the intermediate minimum is achieved at $b_{min}=\sqrt{q}$.
As the condition $\frac{1}{n}\omega > \omega_0(\phi; b)$ is relevant 
only for $b< \sqrt{q}$ whereas for $b> \sqrt{q}$ one gets the 
condition $\frac{1}{n}\omega>r+s$ and
as one has $\omega_0(\phi, b_{min})= r+s$ there is a smooth transition 
in the conditions; and, all in all, $\frac{1}{n}\omega\geq r+s$
is a necessary condition for a class to be a DRY class.

Therefore we get the following theorem 

\noindent
{\bf Theorem on DRY classes.} 
{\em For a class $c=\phi\si + \omega \in H^4(X, {\bf Z})$ to be a DRY class
one has the following conditions 
(where $b$ is some $b\in{\bf R}^{>0}$, $b<\sqrt{q}$, and 
$\omega \in H^4(B, {\bf Z})\cong {\bf Z}$):\\
a) \un{necessary and sufficient:}
$\phi-n(\frac{1}{2}+b)c_1$ is ample and $\frac{1}{n}\omega> \omega_0(\phi; b)$\\
b) \un{sufficient:} $\phi-\frac{n}{2}c_1$ is ample and 
$\omega$ sufficiently large\\
c) \un{necessary:} $\phi-\frac{n}{2}c_1$ is ample 
and $\frac{1}{n}\omega\geq r+s$.}

Here part b) follows immediately from a)
as the ample cone of $B$ is an open set.
So in particular the condition on $\omega$ can be fulfilled in any 
bundle construction which contains a (discrete) parameter 
$\mu$ in $\omega$ such that
$\omega$ can become arbitrarily large if $\mu$ runs in its range of values
(this strategy will be used for spectral and extension bundles).

Note that as the notion of $c$ being a DRY class does not involve any $b$
one should compare these conditions for different $b$.
One then realises that as $b$ becomes larger the condition
on $\phi$ becomes stronger and stronger; on the other hand as $b$ 
increases from $0$ to $\sqrt{q}$ (for a fixed $\phi$)
the condition on $\omega$ becomes weaker. 
Note that, assuming that 
$\phi - \frac{{\cal N}}{2}c_1=A+b{\cal N}c_1$ with an ample class $A$ on $B$, 
one has $(\phi - \frac{{\cal N}}{2}c_1)^2>b^2{\cal N}^2c_1^2$, which is 
$q>b^2$. 
So it is enough to use $b$'s in the interval $0<b< \sqrt{q}$ 
as test parameters. That is the set of DRY classes is the union of 
allowed ranges of $\phi$ and $\omega$ for all these $b$.

{\em Remark:} Note that, 
although for $b_{min}$ the condition on $\omega$ is as weak as possible, 
$\phi-n(\frac{1}{2}+b_{min})c_1$ might not be ample (cf.~the example 
of the tangent bundle given below);
nevertheless the ampleness condition on $\phi$ might be satisifed for a $b$
where the bound $\omega\geq \omega_0(\phi; b)$ turns out to be more stringent
(cf.~again the example).\footnote{Note also
the following property of $b_{min}(\phi)$:
the zero class lies in the boundary of the ample cone; so, if the condition
on $\phi$ is considered for this limiting case, one finds
$\phi-\frac{n}{2}c_1=bnc_1$ such that ($\phi$ is proportional to $c_1$ and)
$b=b_{min}(\phi)$.}

\section{Examples for the DRY-Conjecture}

\resetcounter

In this section we give examples of cohomology classes which are of DRY-form
and appear as second Chern classes of stable $SU(n)$ vector bundles.

\subsection{The tangent bundle}

Let us see whether the cohomology class given by $c_2(TX)$ 
is detected by the weak DRY-Conjecture as a Chern class.
For this we apply the Theorem above to see whether $c_2(X)$ is a DRY class.
The minimum of $\omega_0(\phi; b)$ is assumed at $b_{min}=\sqrt{q}=7/2$ but 
one finds $0<b<7/2$ as the allowed range for $b$ ($c_1$ was assumed ample);
furthermore one has 
$\frac{1}{3}\omega_{TX}=\frac{10}{3}c_1^2+4\geq \omega_0(12c_1, b_{min})
=r+s=\frac{47}{12}c_1^2+\frac{1}{2}$ only for $c_1^2\leq 6$, 
but one has in any case to take a smaller $b$ which makes the bound 
$\omega\geq \omega_0(12c_1; b)$ even more stringent. It suffices however to take
$b$ minimally smaller (which is also optimal for the bound
$\frac{1}{n}\omega \geq \omega_0(12 c_1, b)$ as $\omega_0(12 c_1, b)$
becomes minimally greater for $b$ becoming minimally smaller), 
such that one gets $c_1^2<6$ as precise condition for $c_2(TX)$ to be 
a DRY-class; i.e., in these cases the weak
DRY-conjecture is fulfilled (as $c_3(X)=-60c_1^2$ is negative
actually even the (proper) DRY-conjecture is true); by contrast the cases 
$c_1^2\geq 6$ illustrate that being a DRY-class is only a sufficient 
condition for a class to be realised as Chern class of a stable bundle,
but not a necessary one.

\subsection{Spectral bundles}

In case of spectral cover bundles [\ref{FMW}] one has the 
following expression for $\omega$
\beqa\label{omegaspec}
\omega&=& 
(\lambda^2-\frac{1}{4})\frac{n}{2}\phi(\phi-nc_1)-\frac{n^3-n}{24}c_1^2
\eeqa
Here $\phi$ is an effective class in $B$ with $\phi-nc_1$ 
also effective and $\lambda$
is a half-integer satisfying the following conditions: $\lambda$ 
is strictly half-integral for $n$ being odd;
for $n$ even an integral $\lambda$ requires 
$\phi\equiv c_1 \  ({\rm mod}\  2)$ while a strictly 
half-integral $\lambda$ requires $c_1$ even. 
(In addition one has to assume that the linear system $|\phi|$
is base point free\footnote{\label{basepoint}
a base point is a point common to all members of
the system $|\phi|$ of effective divisors which are linearly equivalent to 
the divisor $\phi$ (note that on $B$ the cohomology class $\phi$ specifies
uniquely a divisor class); on $B$ a Hirzebruch surface ${\bf F_k}$
with base ${\bf P^1}$ $b$ and fibre ${\bf P^1}$ $f$ this amounts to 
$\phi \cdot b\geq 0$}.) 

Often one assumes, 
as we will do here, that $\phi-nc_1$ 
is not only effective but even ample in $B$. 
Then equ.~(\ref{rho - a c1})
shows that we can take $b=1/2$ as upper bound on $a$. 

One has now to check whether the three conditions on $a^2$ 
given above can be fulfilled. 
According to part b) of the theorem in section 2 one learns that 
this is the case
as $\omega$ increases to arbitrarily large values 
when the parameter $\lambda$ is increasing. 

\noindent
{\bf Theorem.}
{\em i) On $X$ an elliptic Calabi-Yau threefold the class 
$c_2(V)=c=\phi \si + \omega$
for $V$ a spectral bundle (of discrete bundle parameters 
$\eta\in H^2(B, {\bf Z})$ and $\lambda \in \frac{1}{2}{\bf Z}$) 
satisfies the assumptions of the weak DRY-Conjecture 
on $c$ for all but finitely many values of the parameter $\la$.\\
ii) For the infinitely many classes $c\in H^4(X, {\bf Z})$
described in i) the {\em weak} DRY-Conjecture is true.\\
iii) For the classes in ii) with negative $\lambda$ the ({\em proper}) 
DRY-Conjecture is true.}

Here part ii) follows of course just from reversing the perspective:
whereas in part i) one started from a given spectral bundle $V$ and 
found a condition ($\la^2$ sufficiently large) that its $c_2(V)$ fulfills
the assumption of the weak DRY-Conjecture, one then turns around the perspective
in part ii), where one has trivially confirmed the existence 
of a stable bundle for a $c=c_2(V)$ which satisfies the assumptions of
the weak DRY-Conjecture.

Part iii) follows as $c_3(V)=2\lambda \phi(\phi-nc_1)$ is
negative for $\lambda$ negative as $\phi\neq 0$ is effective and
$\phi-nc_1$ was assumed ample, so $\phi(\phi-nc_1)$ is positive 
(this argument underlies of course already part ii) as well).

\subsection{Extension bundles}

Stable vector bundles built as an extension
of given stable bundles have been constructed on elliptic Calabi-Yau 
threefolds in [\ref{AC1}]. 
Let $E$ be a rank $r$ $H_B$-stable vector bundle on the base $B$ 
of the Calabi-Yau space with Chern classes $c_1(E)=0$ and $c_2(E)=k$. 
The pullback bundle $\pi^*E$ is then shown to be stable on $X$ 
with respect to the ample class $J=z\sigma +H_B$ 
where $H_B=hc_1$ (with $h\in {\bf R}^{>0}$) [\ref{AC1}]. 
The bundle extension
\beqa
0\to \pi^*E\otimes {\mathcal O}_X(-D)\to V \to {\mathcal O}_X(rD)\to 0
\eeqa
with $D=x\sigma+\alpha$ defines a stable rank $n=r+1$ vector bundle 
if the numerical condition 
equ. (\ref{nonsplit}) is satisfied. We consider here the case $x=-1$ 
for simplicity. For this bundle $c=\phi\sigma+\omega$ is given by 
\beqa
\phi=(n-1)\frac{n}{2}(2\alpha+c_1),\ \ \ \   \omega=k-(n-1)\frac{n}{2}\alpha^2
\eeqa 
As in the spectral case one now has to check whether the three conditions 
on $a^2$ 
given in section 2 can be fulfilled. 
This is the case according to part b) of the theorem in section 2 
if $\alpha$ is chosen such that $2(n-1)\alpha +(n-2)c_1$ is ample and $k$ 
is chosen sufficiently large. Note that this is in agreement 
with the condition that the extension can be chosen nonsplit if
\beqa\label{nonsplit}
\frac{n-1}{2}\Big[n^2\Big(\alpha(\alpha+c_1)
+\frac{c_1^2}{3}\Big)-c_1(2\alpha+\frac{c_1}{3})+1\Big]-k<0
\eeqa
As above in the spectral bundle case we get here the following result.

{\bf Theorem.}
{\em i) On $X$ an elliptic Calabi-Yau threefold the class 
$c_2(V)=c=\phi \si + \omega$
for $V$ an extension bundle (of discrete bundle parameters 
$\alpha\in H^2(B, {\bf Z})$ and $k \in {\bf Z}$) 
satisfies the assumptions of the weak DRY-Conjecture 
on $c$ for all but finitely many values of the parameter $k$.\\
ii) For the infinitely many classes $c\in H^4(X, {\bf Z})$ described in i) 
the {\em weak} DRY-Conjecture is true.\\
iii) For infinitely many classes $c\in H^4(X, {\bf Z})$ 
the ({\em proper}) DRY-Conjecture is true.}

As above section 3.2, part ii) follows from reversing the perspective:
whereas in part i) one started from a given extension bundle $V$ and 
found a condition ($k$ sufficiently large) that its $c_2(V)$ fullfills
the assumption of the weak DRY-Conjecture, one then turns around the perspective
in part ii), where one has now trivially confirmed the existence 
of a stable bundle for a $c=c_2(V)$ which satisfies the assumptions of
the weak DRY-Conjecture.

Part iii) follows as $c_3(V)=
-\frac{(n-1)(n-2)}{3}(c_1^2+3\alpha(\alpha+c_1))-2k<0$ for $k$ 
sufficiently large.

\subsection{A further Example}

Let us finally come back to the motivating question from the introduction.
We will take a stable bundle in the visible sector $V_{vis}$ of the 
heterotic string 
and want to supplement this by a stable bundle $V_{hid}$ of rank $n_h$ 
such that 
the anomaly condition $c_2(V_{vis})+c_2(V_{hid})=c_2(X)$ is satisfied. 
To assure the existence of $V_{hid}$ we will assume the weak DRY conjecture.
So, concretely we will check whether $c:=c_2(X)-c_2(V_{vis})$ is a DRY class. 

Let us take $V_{vis}=\pi^*E$ where $E$ on $B$ is a bundle with $c_2(E)=k$, 
stable with respect to the ample class $H_B$ on $B$. Thus in this case we have 
\beqa
\phi=12c_1, \ \ \ \omega=10c_1^2+12-k
\eeqa
and furthermore one gets the explicit expression for the bound
\beqa
\omega_0=\Big[\frac{6}{n_h}+\frac{1}{6}+\frac{b}{4}
+\frac{(12-\frac{n_h}{2})^2}{4bn_h^2}\Big]c_1^2+\frac{1}{2}.
\eeqa
Let us consider part a) of the theorem of section 2. We get 
$12-n_h(\frac{1}{2}+b)>0$ from the ampleness condition 
(so we are in the regime $b<\sqrt{q}=\frac{12-\frac{n_h}{2}}{n_h}$)
on $\phi$ and $\frac{1}{n_h}\omega\geq \omega_0$ as further condition. 
Note further that the DRY conjecture does not specify a polarization 
with respect to which $V_{hid}$ will be stable; 
so, to get a polystable bundle in total, $V_{vis}$ should be 
stable with respect to an arbitrary ample class; this is true 
in our case $V_{vis}=\pi^*E$ only for $B={\bf P^2}$ 
(where $H^{1,1}(B)$ is onedimensional)
according to Lemma 5.1 of [\ref{AC1}].
This restriction is however in contradiction with 
the necessary condition $\frac{1}{n_h}\omega\geq r+s$ from which
one finds $c_1^2\leq \frac{12-k-n_h/2}{2-n_h/12}$.
Thus, for this (rather special) example of $V_{vis}$ one does not
succeed in complementing (in the sense of satisfying the anomaly equation)
$V_{vis}$ by a hidden bundle. In many more relevant examples for $V_{vis}$,
however, this strategy succeedes [\ref{newdry}].
\\ \\
B.~A. thanks SFB 647 for support. G.~C. thanks the DFG for support 
in the project CU 191/1-1 and SFB 647 and the FU Berlin
for hospitality. 
\appendix

\section{Ample classes on elliptic Calabi-Yau threefolds}

\resetcounter

Let $H=a\si+\rho\in H^2(X, {\bf R})\cong {\bf R}\si + H^2(B, {\bf R})$
be a class on the elliptic Calabi-Yau threefold $X$. Then one has
{\em if $c_1$ is ample}
\beqa
H\; \mbox{ample} &\Longleftrightarrow & a>0, \; \rho-ac_1\; \mbox{ample}.
\eeqa

Consider first the ``$\Longrightarrow$'' direction: 
one has $a=H\cdot F >0$ according to the Nakai-Moishezon criterion
that $H$ is ample just if $H^3>0, H^2\cdot S>0, H\cdot C>0$ for all
irrducible surfaces $S$ and irreducible curves $C$ in $X$; here this
is applied to the fibre $F$. Furthermore, if $c$ is an irreducible curve
in $B$ one has $(\rho-ac_1)\cdot c=H\cdot c\si >0$; and one also has
$(\rho-ac_1)^2=H^2\cdot \si >0$, such that by the same criterion, applied
now on $B$, indeed the class $\rho-ac_1$ is ample.

Consider now the ``$\Longleftarrow$'' direction: the class of an irreducible 
curve $C$ in $X$ is built from the class $F$ and non-negative linear
combinations of classes of the form $c\si$, where $c$ is now the class
of an irreducible curve in $B$; therefore, turning the previous arguments
around, one ends up indeed with $H\cdot C>0$. The classes of irreducible
surfaces are in a similar way built from $\si$ and the $\pi^*c$; for
$H^2\cdot \si$ one can again turn around the previous argument; this is not
so however for $H^2\cdot \pi^*c=ac(2\rho-ac_1)$; in this case we adopt
the additional assumption that $c_1$ is ample, which implies that $\rho$,
and therefore $2\rho-ac_1$ too, is also ample to get the required conclusion.
Similarly one concludes for $H^3=a[\rho^2+(\rho-ac_1)(2\rho-ac_1)]$.

\section{Examples of one-parameter Calabi-Yau spaces}

\resetcounter

Although we treat in the main body of the paper the case of elliptic
Calabi-Yau spaces $X$ let us briefly comment here on
the simpler case where $X$ is a one-parameter space, i.e., $h^{1,1}(X)=1$.

In this case one has the representations (with $k, t\in {\bf Z}$)
\beqa
c&=&kJ^2\\
c_2(X)&=&tJ^2
\eeqa 
where $J$ is a generating element of $H^2(X, {\bf Z})$; for the ample
class $H$ one has $H=hJ$ with $h\in {\bf R}^{>0}$.

The condition for a class $c$ to have DRY form becomes here
\beqa
k&=&n\Big(h^2+\frac{t}{24}\Big)
\eeqa
This amounts to the condition
\beqa
k&>&n\frac{t}{24}
\eeqa
whereas the necessary Bogomolov inequality $c\cdot J>0$ gives just $k>0$
(for example on the quintic one gets the stronger condition $k>\frac{5}{12}n$).
Note that the second Chern class of the tangent bundle always has DRY-form;
thus for this cohomology class the weak DRY-conjecture is satisfied,
and for negative Euler number even the (proper) DRY-conjecture.

Some examples are provided by the complete intersection spaces
${\bf P^4}(5)$, ${\bf P}^5(2,4)$, ${\bf P^5}(3,3)$, 
${\bf P^6}(2,2,3)$, ${\bf P^7}(2,2,2,2)$
with $t=10,7,6,5,4$ and Euler numbers $-200$, $-176$, $-144$, $-144$, $-128$. 
(similarly one can discuss the one parameter cases 
${\bf P_{2,1,1,1,1}}(6)$, ${\bf P_{4,1,1,1,1}}(8)$, ${\bf P_{5,2,1,1,1}}(10)$).

On the quintic one has some further bundles, occurring in the 
list in [\ref{Douglas}], with $c_2(V)=c_2(X)$
with some of them (the first five examples) 
shown to be stable in [\ref{Brambilla}],
which have the same $t$ as $TX$ and also negative $c_3(V)$;
thus these provide further examples of the weak DRY-conjecture 
and actually even of the (proper) DRY-Conjecture.

Physically one has to demand in addition anomaly cancellation. Thus one gets
then in total the condition
\beqa
\frac{n}{24}t<k\leq t
\eeqa
(note that one has here $k_{hid}>0$ for a potential hidden bundle from the 
Bogomolov inequality). 

For the generation number one gets, in the framework of the assumptions of the DRY 
conjecture, the bound
\beqa
N_{gen}<C\frac{n}{2}\Big(\frac{k}{n}-\frac{t}{24}\Big)^{3/2}.
\eeqa

\section*{References}
\begin{enumerate}

\item
\label{DRY}
M.R.~Douglas, R.~Reinbacher and S.-T.~Yau,
{\em Branes, Bundles and Attractors: Bogomolov and Beyond},
math.AG/0604597.

\item
\label{FMW}
R. Friedman, J. Morgan and E. Witten, {\em Vector Bundles and F-Theory},
hep-th/9701162, Comm. Math. Phys. {\bf 187} (1997) 679.

\item
\label{AC1}
B. Andreas and G. Curio, {\em Stable Bundle Extensions on elliptic 
Calabi-Yau threefold},
J. Geom. Phys. {\bf 57}, 2249-2262, 2007, math.AG/0611762.

\item
\label{AFG}
B. Andreas and M. Garcia-Fernandez, {\em Solution of the Strominger 
System via Stable Bundles on Calabi-Yau threefolds}, arXiv:1008.1018 [math.DG].

\item
\label{Strom}
A.~Strominger, {\em Superstrings with Torsion}, 
Nucl. Phys {\bf B 274} (1986)253.

\item
\label{Douglas}
M.R.~Douglas and C.-G.~Zhou, {\em Chirality Change in String Theory}, 
arXiv:hep-th/0403018, JHEP {\bf 0406} (2004) 014.

\item
\label{Brambilla}
M.C.~Brambilla, 
{\em Semistability of certain bundles on a quintic Calabi-Yau threefold},
arXiv:math/0509599.
 
\item
\label{newdry}
B.~Andreas and G.~Curio, to appear.

\end{enumerate}
\end{document}